# The growth of metastable fcc Fe$_{78}$Ni$_{22}$ thin films on H-Si(100) substrates suitable for focused ion beam direct magnetic patterning.

Jonas Gloss[1], Michal Horký[2], Viola Křižáková[3], Lukáš Flajšman[2], Michael Schmid[1], Michal Urbánek[*,2,3] and Peter Varga[1,2]

[1] *Institute of Applied Physics, TU Wien, 1040 Vienna, Austria*
[2] *CEITEC BUT, Brno University of Technology, Purkyňova 123, 612 00 Brno, Czech Republic*
[3] *Institute of Physical Engineering, Brno University of Technology, Technická 2, 616 69 Brno, Czech Republic*
[*] e-mail: michal.urbanek@ceitec.vutbr.cz

**Abstract**

We have studied the growth of metastable face-centered-cubic, non-magnetic Fe$_{78}$Ni$_{22}$ thin films on silicon substrates. These films undergo a magnetic (paramagnetic to ferromagnetic) and structural (fcc to bcc) phase transformation upon ion beam irradiation and thus can serve as a template for direct writing of magnetic nanostructures by the focused ion beam. So far, these films were prepared only on single crystal Cu(100) substrates. We show that transformable Fe$_{78}$Ni$_{22}$ thin films can also be prepared on a hydrogen-terminated Si(100) with a 130-nm-thick Cu(100) buffer layer. The H-Si(100) substrates can be prepared by hydrofluoric acid etching or by annealing at 1200°C followed by adsorption of atomic hydrogen. The Cu(100) buffer layer and Fe$_{78}$Ni$_{22}$ fcc metastable thin film were deposited by thermal evaporation in an ultra-high vacuum. The films were consequently transformed *in-situ* by 4 keV Ar$^+$ ion irradiation and *ex-situ* by a 30 keV Ga$^+$ focused ion beam, and their magnetic properties were studied by magneto-optical Kerr effect magnetometry. The substitution of expensive copper single crystal substrate by standard silicon wafers dramatically expands application possibilities of metastable paramagnetic thin films for focused ion beam direct magnetic patterning.

## 1. Introduction

Magnetic nanopatterning plays a central role in the development of novel devices and magnetic metamaterials with properties unattainable in bulk systems. The patterning is conventionally achieved by using optical or electron-beam lithography in combination with lift-off and etching procedures. However, these methods do not allow, for example, for continuous spatial changes in magnetic properties. The resulting structures usually have sharp magnetic-nonmagnetic transitions leading to, e.g., highly localized demagnetizing fields, or stochastic distribution of pinning sites that cannot be controlled merely by the fabrication process itself. An alternative to the traditional lithography is a fabrication of magnetic elements by ion beam [1]. The ion beam can modify the magnetic properties of a material by alloying [2], intermixing [3] or by a structural change [4]. We have shown that it is possible to embed ferromagnetic body-centered-cubic (bcc) nanostructures in epitaxial paramagnetic face-centered-cubic (fcc) films with a 1 keV $Ar^+$ ion beam [4, 5]. The system used in these works was an fcc Fe grown on a Cu(100) substrate [6, 7]. The ion-beam induces a structural and magnetic phase change by overcoming the potential barrier between the fcc local minimum and the bcc global minimum and we, therefore, call these fcc films metastable. The metastable Fe/Cu(100) spontaneously transforms to bcc phase at 10 ML [8], but this limit can be shifted to 22 ML by dosing CO [9] and even further by alloying Fe with 22% of Ni [10]. We have shown that it is possible to fabricate magnetic nanostructures in these films by ion-beam irradiation using proximity masks [5] or focused ion beams (FIB) and that by suitably chosen FIB patterning procedures it is possible to tune saturation magnetization and also the magnetic anisotropy of the transformed structures [11].

Metastable $Fe_{78}Ni_{22}$/Cu(100) is an excellent system for a one-step fabrication of magnetic nanostructures, but the copper single crystal is an expensive substrate for use in future applications. Here we show, that it is possible to use also standard Si(100) wafers as a base substrate for this system. Silicon has been the most commonly used material in the semiconductor industry and scientific research for decades, not only for its electronic properties but also because of the well-known processes for preparation of very well defined substrates with almost perfect crystallographic properties. Our experiments indicate that the Fe does not grow epitaxially on Si(100), but it is possible to grow a Cu(100) buffer layer on Si(100) [12-14]. The Cu buffer layer grows in the desired (100) crystallographic orientation when the Si(100) surface is unreconstructed, hydrogen-terminated (H-Si), which is most commonly achieved by etching in hydrofluoric acid (HF) [12-17]. The lattice constants of Cu and unreconstructed Si lead to a significant (33.5%) lattice mismatch; however, rotation by 45° decreases the lattice mismatch to 6%.

Additionally to the known HF etching procedure, we have also used an ultrahigh vacuum (UHV) alternative, which requires annealing of Si at 1200°C followed by adsorption of atomic hydrogen. The growth of the 130 nm thick Cu(100) buffer layer on H-Si substrates prepared by both procedures was tested together with the subsequent deposition of the metastable fcc $Fe_{78}Ni_{22}$ films. The transformation to the bcc phase was performed both, *in-situ* by a weakly focused ion beam and *ex-situ* by the FIB. Measurement of the magnetic properties of the transformed thin films and nanostructures performed by Kerr magnetometry show that $Fe_{78}Ni_{22}$/Cu(100)/H-Si(100) is a viable alternative to systems prepared on Cu(100) single-crystal substrates.

## 2. Experimental

The UHV system used for the experiments has a base pressure of $7\times10^{-11}$ mbar (measured by a Bayard-Alpert ionization gauge) and is equipped with a three-pocket e-beam evaporator (Focus EFM3T), a single pocket e-beam evaporator (Focus EFM3) and a Knudsen effusion cell (CreaTec). The samples can be cooled by liquid nitrogen to 100 K and heated by e-beam bombardment of the sample holder plate to 1000 K. Auger electron spectroscopy (AES) was used to check the cleanliness of the substrates and the composition of grown films. The AES spectra have been normalized to the minimum of the average peak-to-peak height of the dominant peak. Presented concentrations correspond to quantitative analysis using relative elemental sensitivity factors [18]. The structure of the surfaces was measured by Low Energy Electron Diffraction (LEED). The LEED images were post-processed with a dark-field subtraction, flat-field normalization [19] and inverted.

The experiments were performed on B-doped (p-type) Si with a resistivity of 5-20 Ωcm. The dimensions of the samples were $3\times12\times0.4$ mm$^3$. Two procedures were applied to obtaining a H-terminated surface. The chemically treated Si was etched for 2.5 minutes in 10% HF to remove the native oxide and to terminate the Si with hydrogen [17], then rinsed for 1 minute in high-purity (Merck Milli-Q) water, dried with argon gas and transferred into a loadlock connected to the UHV chamber within 10 minutes after etching. The H-terminated Si should be inert to the ambient atmosphere for such a period [17]. Then, the sample was outgassed in the UHV chamber at 100°C for 30 minutes. After this procedure, the sample cleanliness was checked by AES and the surface reconstruction by LEED.

In the UHV procedure, Si samples were heated by direct current (DC) heating in a home-built heating stage. Target holders were made of Mo; their design was based on the Omicron sample plates for DC heating. After introducing the samples into UHV, we outgassed the sample holder by heating it to 600 °C for approx. 20h. We then heated the samples by DC to 600 °C for approx. 20 h until the base pressure in the chamber was restored. After the outgassing phase, we annealed the samples repeatedly at 1200 °C by DC for 5 seconds with a 5-second ramp from the outgassing temperature. The highest pressure during the last annealing step was kept below $5\times10^{-10}$ mbar. With this approach, we were able to completely remove both the native oxide and also any organic impurities. The H termination was achieved by a home-built H-cracker based on the design of Bischler [20]. A tungsten capillary with a 0.6 mm inner diameter was heated by 1 keV e-beam bombardment to approx. 1800 °C, which completely dissociated the $H_2$ flowing through it [21]. The end of the capillary was approx. 3 cm from the sample and a liquid-$N_2$-cooled Cu plate between the W tube and the sample limited the sample heating to approx. 1 °C/min. The sample was exposed to atomic hydrogen ($1\times10^{-6}$ mbar $H_2$ backpressure) for 7 mins to achieve a complete H termination. Again, the sample cleanliness was checked by AES and the surface reconstruction by LEED.

Cu was evaporated from two sources, the EFM3, and the effusion cell. The material of the effusion cell crucible contained a small amount of Ca contamination (confirmed by Secondary Ion Mass Spectrometry) which was detectable also in the deposited layer and which turned out to be an essential surfactant needed to stabilize the growth of the Cu buffer layer in the required (100) orientation on UHV treated samples. The temperature of the sample increased by 10 °C via radiation heating during the deposition from the effusion cell. The pressure during the deposition was $1\times10^{-10}$ mbar (with the help of a liquid-$N_2$-cooled cryo baffle and a titanium sublimation pump). Deposition rates were calibrated with a quartz crystal microbalance at the position of the substrate, and the deposition was done at room temperature (RT) unless mentioned otherwise. The deposition rate of both Cu evaporators was 0.06 Ås$^{-1}$ (approx. 5.8 h for 130 nm). The $Fe_{78}Ni_{22}$ layers were evaporated by the EFM3T from a rod with a 2 mm diameter (MaTecK). A repelling voltage of +1.5 kV was applied to a cylindrical electrode (flux monitor) in the orifice of the evaporator to suppress high-energy ions, which may modify the growth mode of the films [22]. The base pressure during the deposition was $8\times10^{-11}$ mbar, which was artificially increased to $5\times10^{-10}$ mbar of CO to stabilize the fcc phase in line with previous observations [10]. The deposition rate of $Fe_{78}Ni_{22}$ was 0.02 Ås$^{-1}$ (approx. 1h for 8 nm). After each deposition step, the surface composition and the crystallographic structure were measured by AES and by LEED (respectively).

Large-Area irradiation of one set of samples was performed *in-situ* by an ion gun (Specs) equipped with a Wien filter by scanning the sample with a time-averaged ion flux of approx. $10^{13}$ cm$^{-2}$s$^{-1}$. To maximize the transformation rate, we used 4 keV Ar$^+$ ions, which penetrate the whole Fe$_{78}$Ni$_{22}$ layer at perpendicular incidence and do not cause significant Fe-Cu intermixing [23]. During the ion beam irradiation, we periodically measured magnetic hysteresis loops by a home-built Surface Magneto-Optical Kerr Effect (SMOKE) apparatus. The SMOKE experiment can be performed in longitudinal or polar geometry (angle of incidence 60° or 30°, respectively, with a spot size of approx. 1 mm). The plane of incidence and the direction of the magnetic field were parallel to the (010) plane of the Si substrate.

To study the potential of the metastable films for magnetic micro- and nanopatterning, we performed a local ion-beam-induced transformation in a high vacuum (approx. 10$^{-7}$ mbar) chamber of a focused ion beam – scanning electron microscope (FIB-SEM) system (Tescan LYRA3). A series of 3×3 µm$^2$ squares was irradiated with the Ga$^+$ FIB at 30 keV with a spot size of 20 nm and a beam current of 40 pA with an increasing ion dose. We varied the irradiation dose between individual areas and then measured the patterns by SEM, Kerr microscopy and micro-focused Kerr magnetometry [24].

## 3. Results and discussion

### a. Surface structure and stoichiometry

The chemically prepared Si(100) samples had a (1×1) diffraction pattern after introduction into the UHV and mild annealing, which is shown by a green square in Figure 1a). The unetched, UHV treated Si(100) showed (2×1)-reconstructed domains after annealing to 1200 °C [blue rectangles in Figure 1d)] which changed into a (1×1) after termination with atomic H [green square in Figure 1e)]. The change in reconstruction confirms that the atomic H had saturated all the Si dangling bonds [25, 26]. The (1×1) diffraction pattern, therefore, means that both approaches are potentially suitable for the growth of epitaxial Cu(100).

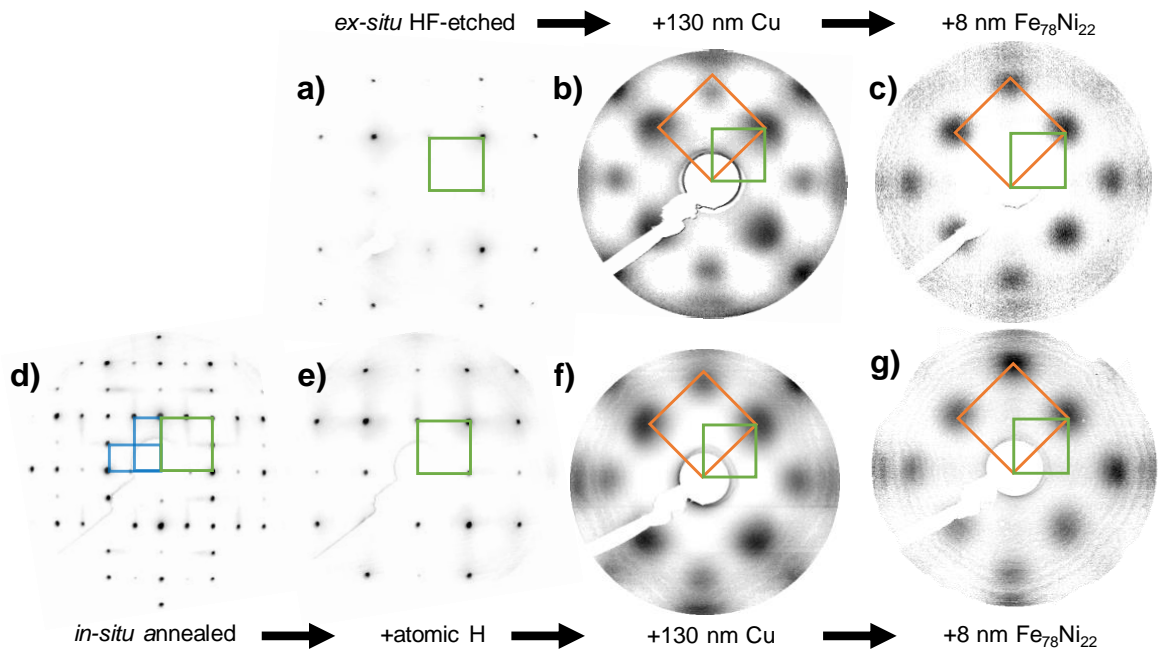

*Figure 1: LEED images taken at 130 eV except for c) and g) which were taken at 120 eV. a) H-Si(100) with its (1×1) unit cell (green in all images) prepared chemically. b) 130 nm Cu/H-Si(100) with its (1×1) unit cell (orange in all images). c) 8 nm $Fe_{78}Ni_{22}$/130 nm Cu/H-Si(100) with its (1×1) unit cell. d) UHV treated Si; blue rectangles are the (2×1) reconstruction, which change into the (1×1) after H-termination, shown in e). Spectra in f) and g) correspond to the deposition steps b) and c) on the UHV treated substrate.*

The AES measurement shows that the HF-etched surface after mild annealing (100 °C) has 15% of C and 2% of O [blue line in Figure 2a),c)]. Because etching in HF for 2.5 minutes completely removes the $SiO_2$, we contribute this signal to imperfect H termination of Si and contamination during transport from the ambient atmosphere before introducing the sample into the UHV, which was impossible to desorb during the mild annealing. Multiple steps of cleaning before the last HF-etch can minimize this contamination [17, 27]. The excessive contamination of the etched Si can prevent the growth of an epitaxial Cu(100) buffer layer; Cu then grows in (111) orientation [28]. H-Si prepared in UHV is much cleaner than the chemically treated Si [compare orange and blue lines in Figure 2a),c)]. The outgassed Si with a native $SiO_2$ is still slightly contaminated by C (black lines in Figure 2), and after annealing, we observe a complete removal of the native oxide and carbon contamination (compare black and red lines in Figure 2). The H termination does not introduce any impurities (orange lines in Figure 2).

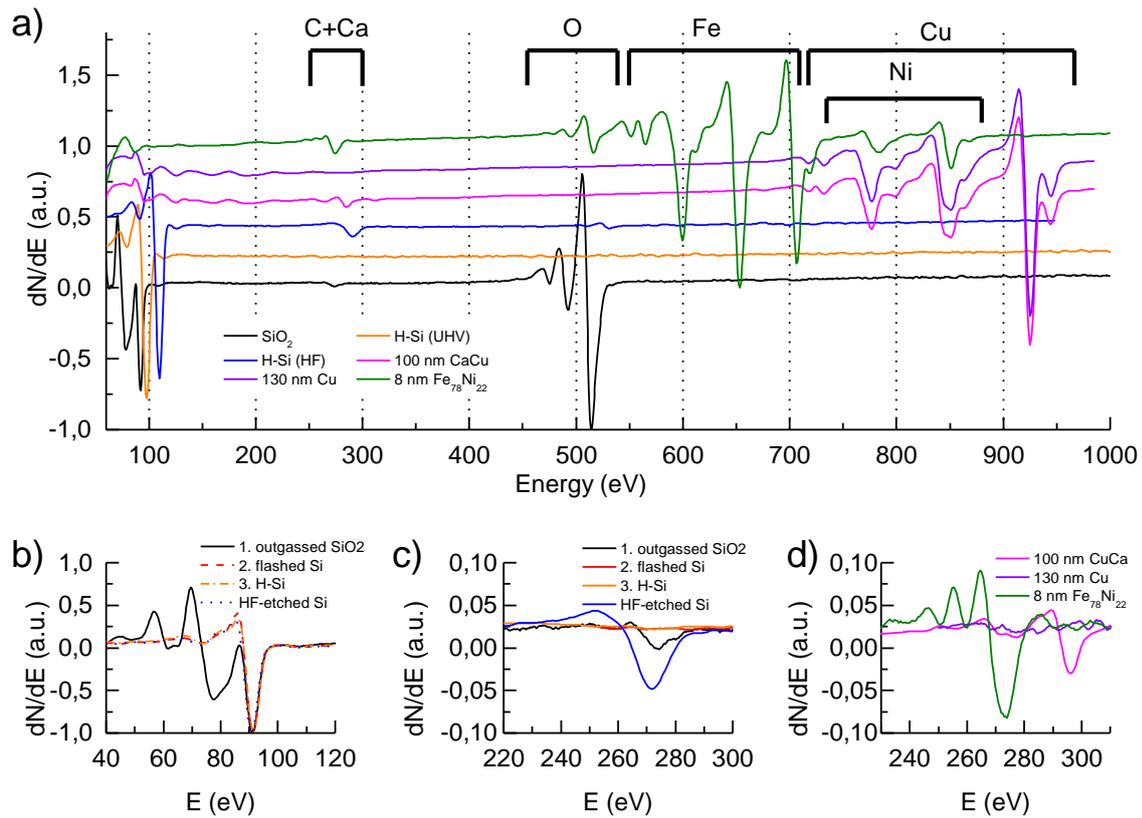

*Figure 2: a) Overview AES of the cleaning (black, orange and blue lines) and deposition (magenta, violet and green lines) steps. The details in b) and c) portray the Si and C peak, respectively. The in-situ-treated Si was outgassed (black), annealed (red) and H-terminated (orange). The blue line corresponds to chemically treated (HF-etched) Si. AES in d) marks the details of C and Ca peaks measured after the deposition of Ca-containing Cu (magenta), Cu (violet) and $Fe_{78}Ni_{22}$ (green), respectively.*

We deposited Cu on both H-Si(100) substrates prepared chemically and in UHV. We grew 130-nm thick films because at this thickness we found a right balance between the film morphology [13] and the deposition time, which was with our evaporators below 6 hours. The deposition of Cu resulted in LEED patterns which confirm that the as-grown Cu film on the chemically treated substrate was very similar to the one on the H-Si prepared in UHV [Figure 1b),f)]. Both films showed a 45° rotation of their unit cell with respect to the underlying substrate and a unit cell reduced by a factor of √2 [marked by orange and green squares in Figures 1b),f)]. The fcc (100) diffraction spots are diffused, and we could also observe their splitting. Both the diffusion and splitting of the diffraction spots lead to a conclusion that the Cu grows as epitaxial islands with terraces wide only a few nm, as already shown by Mewes et al. [14]. Deposition assisted with ion bombardment from the sputter gun did not bring any significant difference in the as-grown films. Deposition at lower or higher temperatures (-20 °C, > 50 °C) led to a growth of a polycrystalline fcc (111) film, in line with previous observations [29, 30]. Post-annealing did not lead to flattening of the film; in fact, we could observe a signal from Si in AES while heating to 150 °C for two hours, in contrast to Lukaszew et al. [31]. The Si signal in AES after post-anneal showed a splitting in energy, which corresponds to the formation of bulk copper silicides, i.e., hybridization between the Si 3p states and the Cu 3d states [16].

The 130-nm Cu buffer layer on the chemically prepared H-Si was completely clean according to AES [violet lines in Figure 2a),d)], indicating that the impurities have either desorbed or got buried by Cu deposition. We were not able to deposit an fcc (100) Cu buffer layer on the UHV prepared H-Si by depositing pure Cu, but we were successful when we used a Ca-containing Mo crucible in the effusion cell. The surface of the 100 nm Cu films deposited from the Ca-containing Mo crucible had 2% of Ca. This is not necessarily concentration in the films; Ca is virtually insoluble in Cu [32], so it is likely that Ca floats to the top during growth and only a negligible part is incorporated. Ca has an fcc structure and a 2.8% lattice mismatch to Si(100) and in this case, it most probably served as a surfactant supporting the epitaxial growth. After deposition of 100 nm fcc (100) Cu [magenta lines in Figure 2a),d)], we have continued with deposition of 30 nm of pure Cu, and AES measurements after the deposition [identical with the violet lines in Figure 2a),d)] revealed that the completed film does not have any Ca on the surface.

In the last step of the film preparation, we deposited 8 nm of $Fe_{78}Ni_{22}$(100) on the epitaxial Cu(100) buffer layer grown on both substrates. Diffraction spots in Figure 1c),g) are commensurate, not split and similarly diffuse as the ones on the Cu buffer layer [Figure 1b),f)]. The spot shape indicates that the $Fe_{78}Ni_{22}$ and the buffer layers beneath have similar morphology and that the structure of the metastable film is also fcc (100). As the films were deposited with a CO background pressure, the surface of the metastable films had 8% C and 5% O measured by AES, as shown by the green line in Figure 2a) and by the detail of the C-peak in Figure 2d). The C and O peaks arise from the dissociation of the CO in which the films are deposited, as described elsewhere [9]. The combination of LEED and

AES demonstrates that it is possible to grow epitaxial films of $Fe_{78}Ni_{22}(100)$ on both H-Si(100) prepared by wet chemistry or in UHV.

### b. *In-situ* magnetic transformation by broad beam irradiation

The experiments presented in Figures 3 and 4 were performed on samples prepared by the UHV treatment; the samples treated by wet chemistry showed equivalent results. SMOKE measurements of the as-deposited films [black circles in Figure 3a)] show a non-zero magnetic signal suggesting a small fraction of the film is already in the bcc ferromagnetic (FM) phase. Irradiation with a dose of $2\times10^{15}$ $cm^{-2}$ leads to partial magnetic transformation; the hysteresis loop [green line in Figure 3 a)] exhibits high coercivity and the Kerr signal reaches approx. 50% of the maximum value measured for the fully transformed film. The film is completely transformed upon applying a transformation dose of $6\times10^{15}$ $cm^{-2}$ [red rectangles in Figure 3a)].

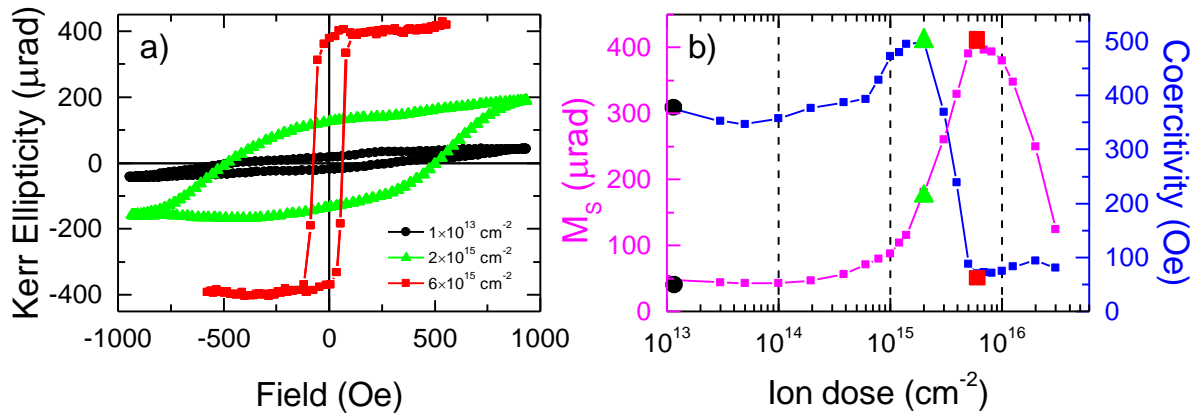

*Figure 3: SMOKE measurements of a metastable 8 nm $Fe_{78}Ni_{22}$ film on 130 nm Cu/H-Si(100). a) the as-deposited film is weakly FM (black). The film partially transforms to bcc (green) after irradiation with dose of $2\times10^{15}$ $cm^{-2}$. The magnetization reaches a maximum at an ion dose $6\times10^{15}$ $cm^{-2}$ (red). b) dependence of the Kerr ellipticity at magnetic saturation ($M_S$, magenta) and coercivity (blue) on the ion dose. Black, green and red symbols correspond to the measurements in (a).*

We demonstrate the development of the magnetic transformation with an increasing ion dose by the magenta line in Figure 3b). We do not detect any measurable signal up to the ion dose of $2\times10^{14}$ cm$^{-2}$. The magnetic signal then increases with the increasing ion dose up to the dose of $6\times10^{15}$ cm$^{-2}$, where it reaches a maximum and then decreases as the magnetic thin film is sputtered away and intermixes with the Cu layer below. The ion dose at which we reach the maximum transformation is equal to the transformation dose for films grown on a Cu(100) single crystal [10]. The Kerr ellipticity of the transformed film is 10% lower than in the case of the films deposited on a Cu(100) single crystal[10], which we attribute to the corrugation of the metastable film arising from the underlying Cu buffer layer.

The coercivity [blue line in Figure 3b)] starts increasing at the same ion dose as the magnetic saturation (around $2\times10^{14}$ cm$^{-2}$) to reach a maximum at $2\times10^{15}$ cm$^{-2}$. It then falls rapidly to its minimum, which matches the maximum of the magnetic saturation at $6\times10^{15}$ cm$^{-2}$. From observations on Cu(100) single crystal substrate we know that increasing the ion dose increases the number and size of bcc nuclei randomly dispersed in the fcc layer. The maximum in the coercive field corresponds to the nuclei reaching the maximum size for single magnetic domains. This property is well known from magnetic nanoparticle studies [33]. Further lowering of the coercive field is attributed to multi-domain states of magnetic particles and more efficient interaction in between the nanoparticles through the stray field of individual particles.

### c. Focused ion beam magnetic patterning

Figure 4 shows the results of patterning the metastable films with the FIB. The grey level in SEM (inset at the bottom of Figure 4) allows for distinguishing the as-deposited (grey) areas and the irradiated (darker or brighter) areas and serves as a valuable pre-characterization tool. The irradiated areas become darker at a dose of $5\times10^{13}$ cm$^{-2}$ but brighten again at doses above $10^{16}$ cm$^{-2}$. The reason for the decrease of the SEM signal is the fcc → bcc structural change, which affects electron channeling and associated secondary electron emission [11]. The last square then shows the signal from the Cu buffer layer after a complete removal of the Fe$_{78}$Ni$_{22}$ film.

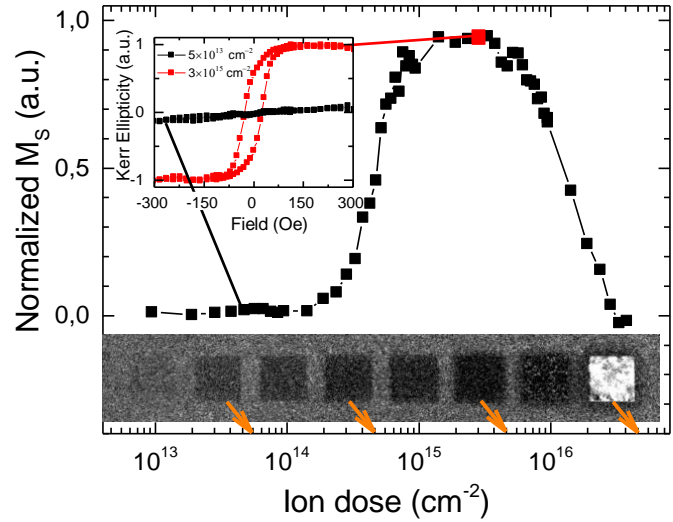

*Figure 4: SEM and Kerr magnetometry analysis of 3×3 µm² squares in the 8-nm $Fe_{78}Ni_{22}$/130 nm Cu/Si(100) irradiated by a 30 keV $Ga^+$ FIB for different ion doses. The Kerr magnetometry confirms that there is no magnetic signal in films irradiated with dose below $5×10^{13}$ cm⁻² (inset top - black line). The film has largest magnetic saturation ($M_S$) when irradiated by $3×10^{15}$ cm⁻² (inset top - red line). The surrounding fcc does not show any hysteresis. The inset at the bottom shows SEM images of the irradiated squares for the ion doses at the abscissa (with a half-an-order step starting at $1×10^{13}$ cm⁻²).*

The data from Kerr magnetometry (Figure 4) show the onset of the magnetic transformation at an ion dose of $1×10^{14}$ cm⁻². The maximum of the magnetization is achieved at ion doses $1×10^{15}$ cm⁻²– $3×10^{15}$ cm⁻². The transformation by 30 keV $Ga^+$ ions qualitatively shows the same behavior as the transformation performed *in-situ* [see Figure 3b)], but the ion dose needed to achieve full transformation is lower for the $Ga^+$. We attribute this difference to the higher cross section for creation of recoils with high enough energy to induce the thermal spike [4]. For example, according to SRIM [23], $Ga^+$ create 4 times more 1 keV recoils than $Ar^+$ as they travel through the metastable film.

The inset in Figure 4 shows the hysteresis loops of the squares irradiated by low ion dose (black line) and at the maximum magnetization (red line). The squares irradiated by low ion dose do not show any FM signal, which is in contrast with the measurement by SMOKE in Figure 3 (black line). We assume this is because the SMOKE collects signal from a large area, and the small fraction of the film which is already in the bcc ferromagnetic (FM) phase is not recognizable by the Kerr magnetometer.

**Conclusion**

We have shown that it is possible to grow metastable, epitaxial fcc $Fe_{78}Ni_{22}$ on H-Si(100) with a Cu(100) buffer layer, and that using a focused ion beam, we can create magnetic micro- and nanostructures with tuneable magnetization. The lower size limit of such nanostructures is given by the focus of the ion beam and the size of the bcc nuclei [5]. The magnetic properties of the films are comparable to the properties of $Fe_{78}Ni_{22}$ films prepared on Cu(100) single crystal substrates [11]. We have presented two (UHV and chemical) possibilities for the preparation of the H-Si(100) substrate of which the UHV alternative has not been used in the past. We have stabilized the Cu(100) buffer layer on the UHV prepared H-Si(100) by addition of Ca. Future research should focus on creating a flat Cu(100) buffer layer by, e.g., introducing a wetting layer. Also, the role of Ca in the stabilization of the epitaxial growth of Cu(100) warrants further investigations.

Metastable films for ion-beam-induced magnetic transformation present a promising system for fabrication of magnetic metamaterials, such as magnonic crystals. They are an alternative to standard lithography approaches, and the possibility to use a standard substrate such as Si(100) is undoubtedly an essential step towards applications in rapid prototyping of magnetic metamaterials (by using FIB) and in suitability for mass production (by ion irradiation through a mask).

**Acknowledgments**

This research has been financially supported by the joint project of Grant Agency of the Czech Republic (Project No. 15-34632L) and Austrian Science Fund (Project No. I 1937-N20) and by the CEITEC Nano+ project (ID CZ.02.1.01/0.0/0.0/16013/0001728). Part of the work was carried out in CEITEC Nano Research Infrastructure (ID LM2015041, MEYS CR, 2016–2019). L.F. was supported by Brno PhD talent scholarship, and V.K. was supported by Thermo Fischer Scientific student scholarship.